\documentstyle[aps,psfig,multicol]{revtex}

\draft

\begin{document}

\title{Towards the solution of the $C_{\rm P}/C_{\rm A}$ anomaly in 
shell-model calculations of muon capture}

\author{T.\ Siiskonen, J.\ Suhonen}

\address{Department of Physics, University of Jyv\"{a}skyl\"{a},
     P.O.Box 35, FIN-40351 Jyv\"{a}skyl\"{a}, Finland}

\author{M.\ Hjorth-Jensen}

\address{Nordita, Blegdamsvej 17, DK-2100 Copenhagen \O, Denmark}

\maketitle

\begin{abstract}
Recently many authors have performed shell-model calculations of
nuclear matrix elements  
determining the rates of the ordinary muon
capture in light nuclei. 
These calculations have employed well-tested
effective interactions 
in large scale shell-model studies.
For one of the nuclei of interest, namely
$^{28}$Si, there exists recent experimental data which can be used to
deduce the value of the ratio $C_{\rm P}/C_{\rm A}$ by using the calculated
matrix elements. Surprisingly enough, all the abovementioned shell-model results
suggest a very small value ($\simeq 0$) for $C_{\rm P}/C_{\rm A}$, quite
far from the PCAC prediction and recent data on muon capture in
hydrogen. We show that this rather disturbing anomaly is solved by employing
effective transition operators. This finding
is also very important in studies of the scalar coupling of the weak 
charged current of leptons and hadrons.
\end{abstract}


\begin{multicols}{2}

The calculation of the nuclear matrix elements involved in the
ordinary (non-radiative) capture of stopped negative muons 
by atomic nuclei has
been of considerable interest because they enable to access the 
structure of the effective weak baryonic current. Due to the large mass
of the captured muon the process involves a large energy release 
(roughly 100 MeV) and thus surveys the baryonic current deeper than 
ordinary beta decay or electron capture.
In particular, the role of the induced pseudoscalar coupling $C_{\mathrm P}$
becomes prominent. Furthermore, the same nuclear matrix elements can be
used in the context of studies of the fundamental structure of the weak
charged current of leptons and hadrons, in particular concerning
contributions coming from the yet undetected scalar coupling of the current.

In the past there have been many calculations of nuclear matrix 
elements involved in the muon-capture processes. These calculations
have been either very schematic ones \cite{mor,GIL65,PAR78} or more 
realistic ones using various nuclear models 
\cite{GIL65,PAR78,ERI64}. Ultimately, all these
calculations have aimed at predicting the ratio $C_{\rm P}/C_{\rm A}$ 
of the induced pseudoscalar and axial-vector 
coupling strengths of the weak baryonic current by exploiting the
scarce experimental information 
on muon-capture rates in light nuclei.
These calculations seem to suggest wide ranges of values 
($4\le C_{\rm P}/C_{\rm A}\le 39$ for $^{12}$C \cite{mor}, 
$3\le C_{\rm P}/C_{\rm A}\le 20$ for $^{16}$O and $^{28}$Si 
\cite{GIL65,PAR78}, 
$C_{\rm P}/C_{\rm A}\ge 13$ for $^{16}$N \cite{ERI64}), for this ratio.
For reference the nuclear-model independent Goldberger-Treiman value
is $C_{\rm P}/C_{\rm A}=6.8$, 
which is obtained using
the partially conserved axial current hypothesis (PCAC). 
It should be
reasonable to assume that this relation between the induced pseudoscalar
and axial-vector coupling constants will survive in finite nuclei,
although corrections may occur e.g., due to mesonic corrections in the weak
vertices. In particular, this result should be roughly recovered using
nuclear-structure calculations assuming the impulse approximation.
Renormalizations within the impulse approximation have already
been 
extensively discussed for the $C_{\rm A}$ coefficient in the context 
of beta decay and electron capture. On the experimental side
the interval $6.8\le C_{\rm P}/C_{\rm A}\le 10.6$ was obtained in the
ordinary muon capture in hydrogen \cite{bar} and
the interval $8.8\le C_{\rm P}/C_{\rm A}\le 10.8$ was measured at TRIUMF
for the radiative muon capture in hydrogen \cite{jon}.

Recently, also the nuclear shell-model has been used to
calculate the needed nuclear matrix elements for muon capture 
\cite{GMI90,KUZ94,JOH96,MOF97,sii}. 
The calculation of \cite{GMI90} was performed for $^{12}$C in the
$0s$ and $0p$ shells but no definitive conclusions for 
the $C_{\rm P}/C_{\rm A}$ ratio could be reached
on the basis of their computed matrix elements.
In \cite{KUZ94} the muon capture transitions in $^{11}$B and $^{12}$B
were studied but no conclusions were made with respect to the
possible range of the $C_{\rm P}/C_{\rm A}$ value.
In \cite{JOH96} a range of $4.1\le C_{\rm P}/C_{\rm A}\le 8.9$ was obtained
for the ordinary muon capture in $^{23}$Na by fitting sevaral partial
capture rates to the various final states in $^{23}$Ne. In \cite{sii} 
several light nuclei in the $0p$, $1s0d$ and $0p$-$1s0d$ shell-model 
spaces were studied
by applying well-established effective two-body interactions in these spaces.
In Ref.\ \cite{sii} a 
comparison of the calculated partial muon-capture rates 
with the available
data indicated a notable suppression (when compared with the
above mentioned Goldberger-Treiman value $C_{\rm P}/C_{\rm A}=6.8$)
for $C_{\rm P}/C_{\rm A}$ ($-3.0\le C_{\rm P}/C_{\rm A}\le 2.5$) 
in the case of $^{12}$C but nothing definitive
(the interval $-7\le C_{\rm P}/C_{\rm A}\le 15$ was examined) 
could be said in case of the other studied nuclei ($^{16}$O,$^{20}$Ne,
$^{23}$Na, $^{28}$Si and $^{32}$S) due to controversial information
coming from different partial capture rates in the same nucleus (no best
fit procedure was attempted as in \cite{JOH96}).

Very recently two important measurements of correlation coefficients of
$\gamma$-radiation anisotropy in the capture of a polarized negative muon
have been reported \cite{MOF97,bru}. For the allowed muon capture the
angular correlation between the emitted $\gamma$-radiation and the neutrino
is scaled by the coefficient $\alpha$ \cite{PAR78,MOF97} which is related
to the coefficient 
        \begin{equation}
       x\equiv M_1(2)/M_1(-1)
       \label{xdef}
        \end{equation}
of \cite{bru} by
        \begin{equation}
          \alpha = {\sqrt{2}x - x^2/2 \over 1+x^2}\ .
	  \label{alphax}
        \end{equation}
Here the quantities $M_1(2)$ and $M_1(-1)$ are linear combinations of 
the coupling coefficients of the baryonic weak current and
the nuclear matrix elements for muon capture 
[101], [121], [011$p$], [111$p$] in the
notation of \cite{mor} as defined in \cite{bru}. 
Using the Fujii-Primakoff approximation (FPA)
\cite{FUJ59} leads to the 
following nuclear-model independent result \cite{bru} 
        \begin{equation}
          (C_{\rm P}/C_{\rm A})_{\rm FPA} = {68.37x\over x + 
          \sqrt{2}} - 2.71 \ .
        \end{equation}
In the two measurements the following parameter values were extracted:
$x=0.254\pm 0.034$ \cite{bru} and $x=0.315\pm 0.08$ \cite{MOF97}, leading
to the ratios $(C_{\rm P}/C_{\rm A})_{\rm FPA} = 7.7\pm 1.2$ and
$(C_{\rm P}/C_{\rm A})_{\rm FPA} = 9.7\pm 2.6$, respectively, 
compatible with the PCAC prediction. 

Abandoning the FPA and determining the
ratio $C_{\rm P}/C_{\rm A}$ from Eqs. (\ref{xdef}) and 
(\ref{alphax}) using the complicated
expressions of $M_1(2)$ and $M_1(-1)$, leads to rather contradictory results
between different realistic nuclear models. In \cite{bru} the following values
of $C_{\rm P}/C_{\rm A}$ were extracted using the different realistic
nuclear models: $C_{\rm P}/C_{\rm A} = 3.4\pm 1.0$ and
$C_{\rm P}/C_{\rm A} = 2.0\pm 1.6$ for the matrix elements of \cite{CIE76}
and \cite{PAR81}, respectively. In the measurement of \cite{MOF97}
the corresponding extracted values are $C_{\rm P}/C_{\rm A} = 5.3\pm 2.0$ 
\cite{CIE76}
and $C_{\rm P}/C_{\rm A} = 4.2\pm 2.5$ \cite{PAR81}. These results would
indicate a small quenching of the $C_{\rm P}/C_{\rm A}$ ratio with respect
to the PCAC value. However, using the more realistic nuclear matrix elements
obtained from the full $1s0d$ shell-model calculation with the
$sd$-shell effective interaction (USD) 
of Wildenthal \cite{wil}, would lead to the
very contradictory result of $C_{\rm P}/C_{\rm A} = 0.0\pm 3.2$ \cite{MOF97}
by employing the computed matrix elements of Junker {\it et al.} \cite{JUN95}. 
The same situation occurs for the calculation of \cite{sii}, 
verifying
thereby  the
calculation of \cite{JUN95}. This result is rather 
surprising when one takes into account 
the fact that the USD interaction is found to be very 
succesful 
for $1s0d$ nuclei in
reproducing various spectroscopic quantities like energy spectra, 
Gamow-Teller decay properties, electromagnetic moments and transitions 
(see e.g., \cite{car,bro}) as well as strength functions of
charge-exchange reactions \cite{JOH96}. 

The above anomaly in the $C_{\rm P}/C_{\rm A}$ predictions from
state-of-the-art shell-model calculations is rather disturbing when
contrasted with the experimental data.
To give a deeper insight, we investigate 
in the present Letter the capture reaction 
${}^{28}_{14}{\mathrm Si}(0^+_{\mathrm gs})+\mu^-
\rightarrow{}^{28}_{13}{\mathrm Al}(1^+_3)+\nu_\mu$ within
the full shell-model framework and try to evaluate
the ratio $C_{\rm P}/C_{\rm A}$ through the quantity $x$ and its measured
values reviewed above. The needed muon-capture formalism is developed 
in \cite{mor} and reviewed in the case
of shell-model calculations in \cite{KUZ94,sii}. The 
two-body interaction
matrix elements used in the shell-model calculation 
are given by the  
USD interaction \cite{wil} and a microscopic effective
interaction based on the recent 
charge dependent nucleon-nucleon
interaction of Machleidt and co-workers, 
the CD-Bonn interaction \cite{mac96}. This is a meson-exchange 
potential model. Based on this nucleon-nucleon
interaction, we derive a $G$-matrix appropriate for 
the $1s0d$ shell. This $G$-matrix is in turn used in a 
perturbative summation of higher-order terms using the 
so-called $\hat{Q}$-box approach 
described in e.g., Ref.\ \cite{hko95}.
All diagrams through third order in perturbation
theory were used to define the  $\hat{Q}$-box, while folded
diagrams were summed to infinite order, see Ref.\  \cite{hko95}
for further details. Below we will label results obtained with this
effective interaction by CD-Bonn. We have also performed similar 
calculations with the Nijmegen I \cite{nim94} 
potential. The results were very
close to those obtained with the CD-Bonn potential and hence
skipped in the discussion below. 

These effective interactions
are defined within the 
$1s0d$ shell, using $^{16}$O as closed shell core, 
and we perform a full shell-model calculation
using the code OXBASH \cite{oxb}. 

It is important to keep in mind that the USD
interaction is fitted to reproduce several properties of
$1s0d$ shell nuclei, whereas the effective interaction based on 
the CD-Bonn meson-exchange potential model starts from the 
bare nucleon-nucleon interaction. The effects of the 
nuclear medium are then introduced through various terms 
in the many-body expansion.

The single-particle
matrix elements were evaluated in the harmonic-oscillator basis
by numerical integration of the radial
part containing overlap of the initial and final harmonic-oscillator
wave functions and the muonic $s$-state wave function. 
A harmonic oscillator basis was also employed in our calculation
of the $G$-matrix which enters the computation of the effective
interaction, using an oscillator parameter of $1.72$ fm.

In addition to using the
traditional bare transition operators in the
evaluation of the nuclear
matrix elements for muon capture the aim here is
to employ 
effective transition operators as well. Such a calculation
will be referred to as the renormalized one in the discussion
below. To obtain effective one-body 
transition operators for muon capture, we evaluate all 
effective operator diagrams through second-order in the 
$G$-matrix obtained with  the CD-Bonn interaction, 
including folded diagrams. Such diagrams
are discussed in the reviews by Towner \cite{towner87}
and Ellis and Osnes \cite{eo77}.
Terms arising from meson-exchange currents have 
been neglected and we omit
wave function renormalizations in the calculation
of the effective operators.
Intermediate state excitations in each diagram
up to 
$6\hbar\omega$ in oscillator energy were included, as was also the
case for the effective shell-model interaction discussed above.

As such, the determination of both the microscopic 
shell-model effective interaction and the effective
one-body operators is done at the same level of many-body
approach. Further details will be presented elsewhere \cite{ssh98}.
In general, higher order terms introduce a correction of the 
order of 10\% compared with the bare transition operator.
Since the USD interaction is an effective one acting
within the $1s0d$ shell only, it is not possible to calculate
a corresponding effective operator employing perturbative
many-body methods. We will therefore employ the effective
operators  obtained
from the CD-Bonn interaction for the USD calculation as well.

\begin{figure}
       \begin{center}
       {\centering
          \mbox{\psfig{figure=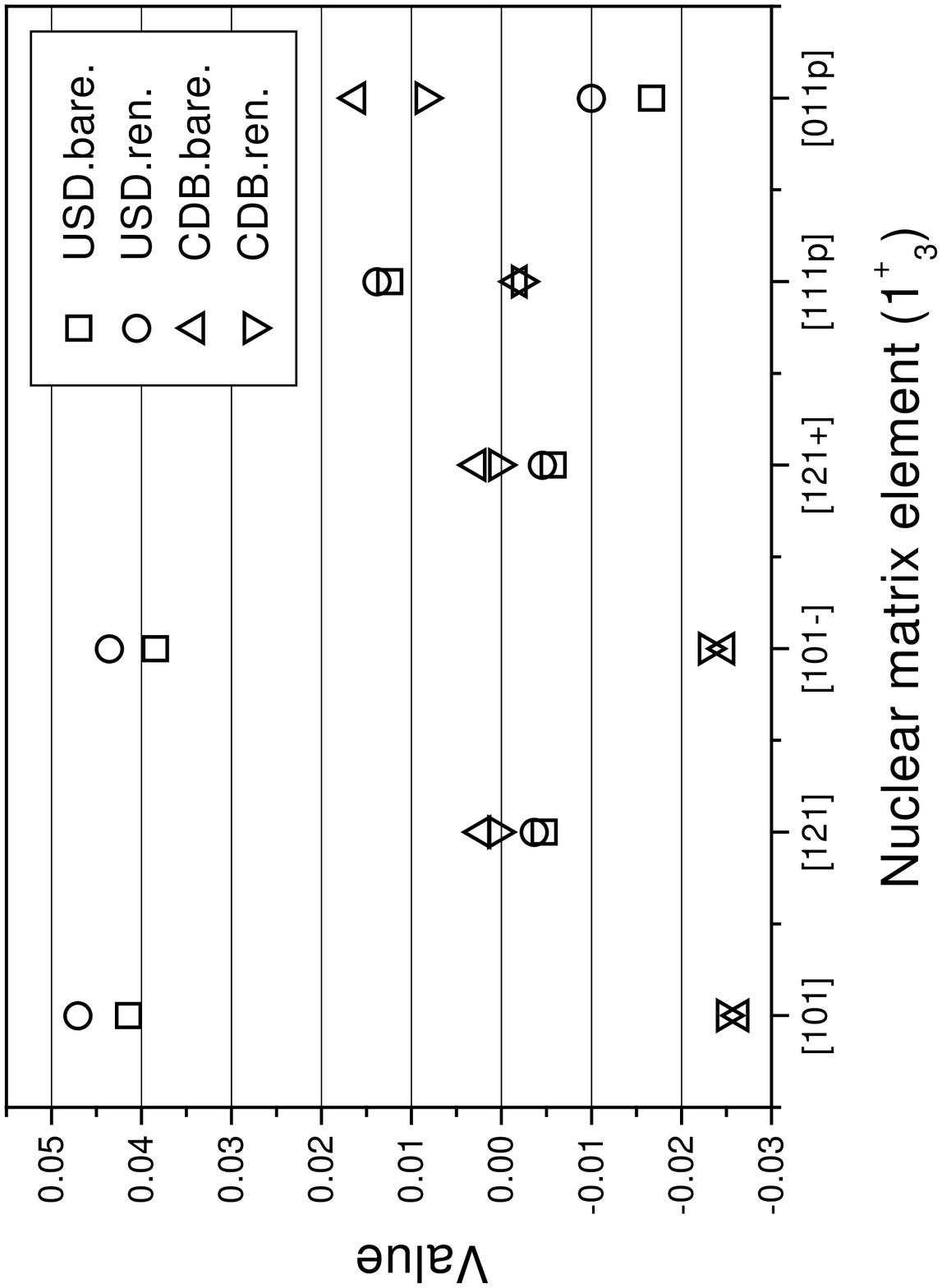,height=20cm,width=15cm}}}
        \caption{The values of all the relevant 
         reduced nuclear matrix 
        elements for muon capture computed using the USD and CD-Bonn interactions with and
        without renormalization of the involved transition operators.
        The recoil matrix elements $[\dots p]$ are
        in units of fm$^{-1}$.}
        \label{nme}
\end{center}
\end{figure}

\begin{figure}
     \begin{center}
      {\centering
      \mbox{\psfig{figure=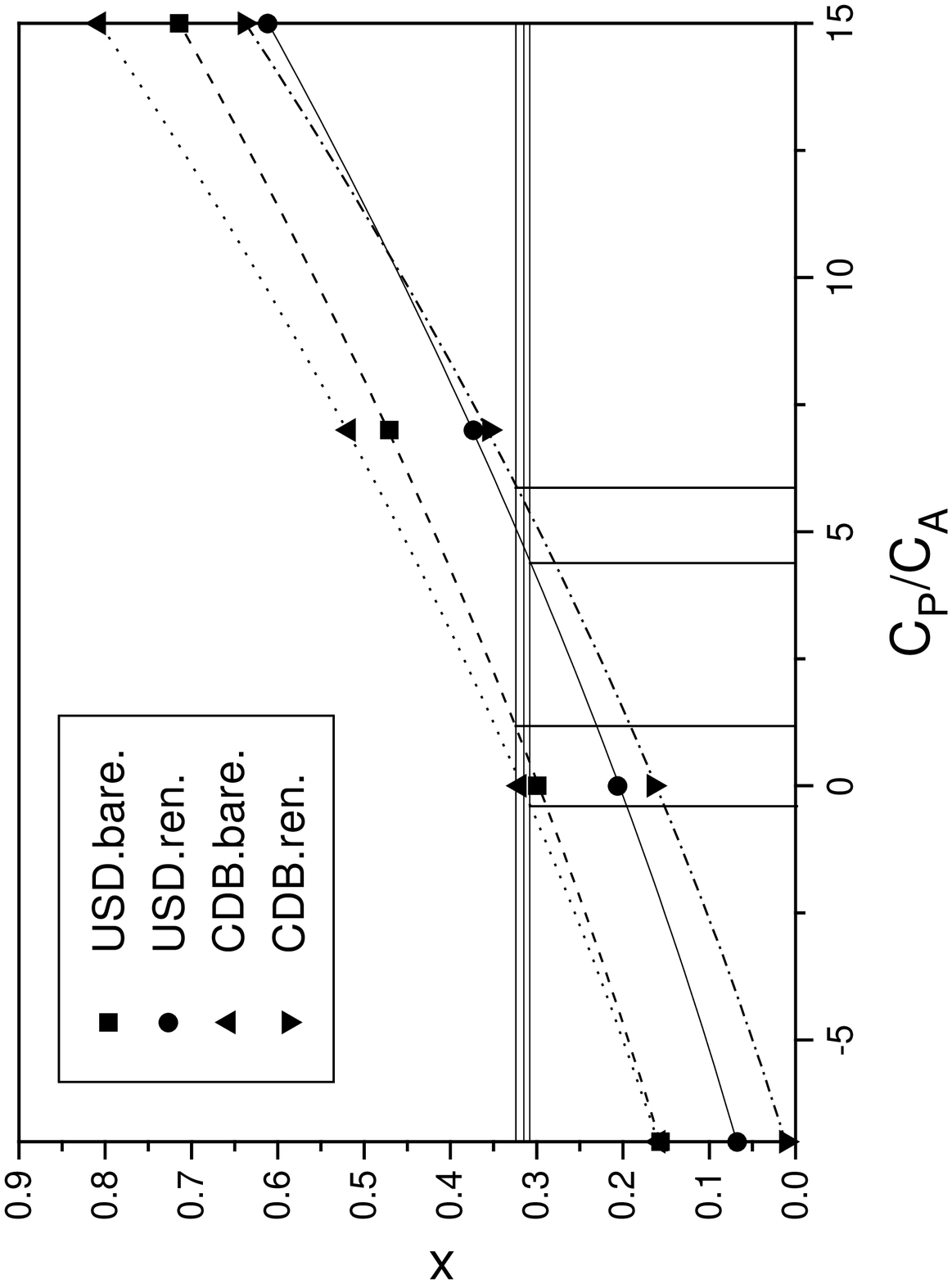,height=20cm,width=15cm}}}
        \caption{The ratio $x=M(2)/M(-1)$ as a function of the ratio 
        $C_{\mathrm P}/C_{\mathrm A}$. The experimental 
        value \protect\cite{MOF97} with the error 
        limits is indicated by the horizontal lines.}
        \label{x}
      \end{center}
\end{figure}

Our results are summarized in Figs.\ \ref{nme} and \ref{x} with
$C_{\mathrm V}/C_{\mathrm A}=-1.0$ (for discussion of aspects concerning
this ratio see \cite{sii} for more details).

Inspection of the nuclear matrix elements for muon
capture of Fig.\ \ref{nme} shows
that in the case of the USD interaction the renormalization 
corrections are largest for the largest matrix elements. 
This, in turn, is reflected in the capture
rates, where the renormalized USD calculation shows much
better agreement with experiment than the bare one. Thus,
the effect of the renormalization is to bridge the gap between the shell-model
calculation and experiment. 
For the CD-Bonn interaction the changes in the reduced matrix
elements are more subtle. Generally the bare matrix 
elements have larger
absolute values than the renormalized ones. 
The USD interaction favours the extreme 
single-particle limit for the 
ground state of ${}^{28}$Si, where all the valence
nucleons lie on the $d_{5/2}$ orbit. The CD-Bonn occupancies are more evenly
distributed over the various orbits, the largest one being $(d_{5/2})^6\,
(s_{1/2})^4\,(d_{3/2})^2$. Also the final
state in ${}^{28}$Al has very different structure: the USD
calculation indicates
that the partition $(d_{5/2})^{10}(s_{1/2})^1(d_{3/2})^1$ should be the most
important part in the $1^+_3$ wave function whereas the CD-Bonn interaction
predicts the state to be mostly of the $(d_{5/2})^8(s_{1/2})^2(d_{3/2})^2$
configuration. In general, CD-Bonn calculations favour the configurations which
have several particles lifted up from the extreme single-particle shell-model
limit which is, on the contrary, favoured by the USD interaction. 

To extract an estimate for the ratio $C_{\rm P}/C_{\rm A}$, we have
plotted $x$ of Eq.\ (\ref{xdef}) as a function of $C_{\rm P}/C_{\rm A}$. The
experimental value $x=0.315\pm0.08$ is taken 
from \cite{MOF97}. With this choice 
we obtain from Fig.\ \ref{x} the range
$-0.4\le C_{\rm P}/C_{\rm A}\le 1.2$ 
for the bare USD and CD-Bonn calculations
in agreement with the USD result of \cite{JUN95} cited in \cite{MOF97}.
Thus, 
for both of the adopted interactions the bare result is almost the
same in spite of the basic difference in the origin 
of the used interactions and differences in the resulting wave 
functions, hinting to a strong
suppression of the $C_{\rm P}/C_{\rm A}$ ratio for the studied
muon-capture transition in the framework of the nuclear shell-model.
As discussed before, this contradicts the
shell-model calculation of the partial capture rates in 
$^{23}$Na (with a fitted $C_{\rm P}/C_{\rm A}$ ratio)  
as well as experimental data on hydrogen.

Also the 
USD and CD-Bonn calculations with renormalized transition
probabilities
agree almost exactly and both yield a very different value for
$C_{\rm P}/C_{\rm A}$ than the calculation using bare operators.
The renormalized result is $4.4\le C_{\rm P}/C_{\rm A}\le 5.9$,
close to the PCAC value and the results of the  
calculations reviewed in \cite{bru} and \cite{MOF97}. This result
encourages us to believe that the anomaly in the $C_{\rm P}/C_{\rm A}$ ratio,
inherent in the sophisticated shell-model calculations, has been lifted
by introducing effective renormalized transitions operators acting in
the muon-capture process. In this way the renormalized shell-model
calculations yield a value for $C_{\rm P}/C_{\rm A}$ in $^{28}$Si 
compatible with data and expectations coming from other theoretical 
approaches.

In conclusion, it is found that the 
renormalization of the transition
operators is essential in the ordinary muon capture
even in the case of an empirical effective interaction, 
which otherwise
reproduces the spectroscopy of the involved nuclei extremely well. 
This is
confirmed by the anisotropy data
through the quantity $x=M_1(2)/M_1(-1)$. The renormalization has helped to
solve the $C_{\rm P}/C_{\rm A}$ problem in the shell-model calculations.
Further analysis of this observation is in progress for other light
nuclei in terms of the capture rates since anisotropy data are only
available for $^{28}$Si. For other nuclei of interest the 
theoretical
situation is more complicated than is the case in 
$^{28}$Si since they are
either situated at the interface of the $0p$ and $1s0d$ shells or 
$1s0d$ and $1p0f$ shells, complicating thereby the evaluation
of an effective interaction and increasing the dimensionality
of the shell-model calculation. Research along such lines is
in progress \cite{ssh98}.\newline   
\newline
J.\ S.\ thanks the Academy of 
Finland. This research has also been supported by
the Nordic Academy of Advanced Studies (NorFA).

\end{multicols}

\clearpage

\clearpage


\begin{thebibliography}{100}
\bibitem{mor} M.\ Morita and A.\ Fujii, Phys.\ Rev.\ {\bf 118}, 606 (1960).
\bibitem{GIL65} V.\ Gillet and D.\ Jenkins, Phys.\ Rev.\ {\bf 140}, B32 (1965).
\bibitem{PAR78} R.\ Parthasarathy and V.\ N.\ Sridhar, Phys.\ Rev.\ 
C {\bf 18}, 1796 (1978).
\bibitem{ERI64} T. Ericson, J. C. Sens and H. P. C. Rood, 
Nuovo Cim. {\bf 24}, 51 (1964).
\bibitem{bar} G.\ Bardin {\it et al.}, Phys.\ Lett.\ {\bf B104}, 320 (1981).
\bibitem{jon} G.\ Jonkmans {\it et al.}, Phys.\ Rev.\ Lett.\ {\bf 77}, 4512 
	(1996).
\bibitem{GMI90} M.\ Gmitro {\it et al.}, Nucl. Phys.\ {\bf A507}, 707 (1990).
\bibitem{KUZ94} V.\ A.\ Kuz'min, A.\ A.\ Ovchinnikova and T.\ V.\ Tetereva, 
Physics of Atomic Nuclei {\bf 57}, 1881 (1994).
\bibitem{JOH96} B.\ L.\ Johnson {\it et al.}, Phys.\ Rev.\ C {\bf 54}, 
2714 (1996).
\bibitem{MOF97} B.\ A.\ Moftah {\it et al.}, Phys.\ Lett.\ {\bf B395}, 
157 (1997).
\bibitem{sii} T.\ Siiskonen, J.\ Suhonen, V.\ A.\ Kuz'min, and T.\ V.\ Tetereva,
        Nucl.\ Phys.\ {\bf A}, in press.
\bibitem{bru} V.\ Brudanin {\it et al.}, Nucl.\ Phys.\ {\bf A587}, 577 (1995).
\bibitem{FUJ59} A.\ Fujii and H.\ Primakoff, Nuovo Cim. {\bf 12}, 327 (1959).
\bibitem{CIE76} S.\ Ciechanowicz, Nucl. Phys.\ {\bf A267}, 472 (1976).
\bibitem{PAR81} R.\ Parthasarathy and V.\ N.\ Sridhar, Phys.\ Rev.\ 
C {\bf 23}, 861 (1981).
\bibitem{wil} B.\ H.\ Wildenthal, Prog.\ Part.\ Nucl.\ Phys.\ {\bf 11}, 5 (1984).
\bibitem{JUN95} K.\ Junker {\it et al.}, Proc. IV Int. Symp. on Weak and
Electromagnetic Interactions in Nuclei (WEIN'95), Osaka, Japan, 1995,
eds. H.\ Ejiri, T.\ Kishimoto, and T.\ Sato (World Scientific, Singapore, 1996) 
p. 394.
\bibitem{car} M.\ Carchidi, B.\ H.\ Wildenthal, and B.\ A.\ Brown, Phys.\ Rev.\
        C {\bf 34}, 2280 (1986).
\bibitem{bro} B.\ A.\ Brown and B.\ H.\ Wildenthal, Nucl.\ Phys.\ {\bf A474}, 290
        (1987).
\bibitem{mac96} R.\ Machleidt, F.\ Sammarruca, and Y.\ Song,
Phys.\ Rev.\ C {\bf 53}, R1483 (1996).
\bibitem{hko95} M.\ Hjorth-Jensen, T.\ T.\ S.\ Kuo, and E.\ Osnes,
Phys.\ Rep.\ {\bf 261}, 125 (1995).
\bibitem{nim94} V.\ G.\ J.\ Stoks, R.\ A.\ M.\ Klomp, 
C.\ P.\ F.\ Terheggen, and J.\ J.\
de Swart, Phys.\ Rev.\ C {\bf 49},   2950 (1994).
\bibitem{oxb} B.\ A.\ Brown, A.\ Etchegoyen. and W.\ D.\ M.\ Rae,
        The computer code OXBASH, MSU-NSCL report 524 (1988).
\bibitem{towner87} I.\ S.\ Towner, Phys.\ Rep.\ {\bf 155}, 263 (1987); B.\ Castel and I.\ S.\ Towner, {\em Modern Theories
of Nuclear Moments}, (Clarendon Press, Oxford, 1990) pp.\ 55. 
\bibitem{eo77} P.\ J.\ Ellis and E.\ Osnes, Rev.\ Mod.\ Phys.\
{\bf 49}, 777 (1977).
\bibitem{ssh98} T.\ Siiskonen, J.\ Suhonen, and M.\ Hjorth-Jensen,
to be published.
\end{thebibliography}
\end{document}